\documentclass[5p,times]{elsarticle}
\usepackage{listings}
\usepackage{amsmath,amssymb,amsfonts}
\usepackage{algorithmic}
\usepackage{graphicx}
\usepackage{textcomp}
\usepackage{xcolor}
\usepackage{mathtools}
\usepackage{xspace}
\usepackage{float}
\usepackage{subfig}

\DeclarePairedDelimiter\floor{\lfloor}{\rfloor}
\DeclarePairedDelimiter\nint{\lfloor}{\rceil}
\def\BibTeX{{\rm B\kern-.05em{\sc i\kern-.025em b}\kern-.08em
    T\kern-.1667em\lower.7ex\hbox{E}\kern-.125emX}}
\usepackage{etoolbox}
\makeatletter
\patchcmd{\@makecaption}
  {\scshape}
  {}
  {}
  {}
\makeatother

\biboptions{sort&compress}

\newcommand{\csr}{CSR-$k$\xspace}
\newcommand{\spmv}{SpMV\xspace}
\newcommand{\cuspmvt}{GPU\-SpMV-3\xspace}
\newcommand{\cuspmvtf}{GPU\-SpMV-3.5\xspace}
\newcommand{\cuspmv}{GPU\-SpMV\xspace}
\newcommand{\cuda}{CUDA\xspace}
\newcommand{\kokkos}{Kokkos\-Kernels\xspace}
\newcommand{\csrt}{CSR-3\xspace}
\newcommand{\csrf}{CSR5\xspace}
\newcommand{\tile}{TileSpMV\xspace}

\newcommand{\cusparse}{cuSPARSE\xspace}
\newcommand{\nvidia}{NVID\-IA\xspace}

\begin{document}
\begin{frontmatter}

\title{Heterogeneous Sparse Matrix-Vector Multiplication via Compressed Sparse Row Format}

\author{Phillip Allen Lane\corref{cor1}\fnref{fn1}}
\ead{phillip.lane@uah.edu}
\cortext[cor1]{Corresponding author}
\fntext[fn1]{Graduate Student}

\author{Joshua Dennis Booth\fnref{fn2}}
\ead{joshua.booth@uah.edu}
\fntext[fn2]{Assistant Professor of Computer Science}

\begin{abstract}
Sparse matrix-vector multiplication (\spmv) is one of the most important kernels in high-performance computing (HPC), yet \spmv normally suffers from ill performance on many devices.
Due to ill performance,  \spmv normally requires special care to store and tune for a given device.
Moreover, HPC is facing heterogeneous hardware containing multiple different compute units, e.g., many-core CPUs and GPUs.
Therefore, an emerging goal has been to produce heterogeneous formats and methods that allow critical kernels, e.g., \spmv, to be executed on different devices with portable performance and minimal changes to format and method.
This paper presents a heterogeneous format based on CSR, named \csr, that can be tuned quickly and outperforms the average performance of Intel MKL on Intel Xeon Platinum 8380 and AMD Epyc 7742 CPUs while still outperforming  \nvidia's \cusparse and Sandia National Laboratories' \kokkos on \nvidia A100 and V100 for regular sparse matrices, i.e., sparse matrices where the number of nonzeros per row has a variance $\leq 10$, such as those commonly generated from two and three-dimensional finite difference and element problems. 
In particular, \csr achieves this with reordering and by grouping rows into a hierarchical structure of super-rows and super-super-rows that are represented by just a few extra arrays of pointers.
Due to its simplicity, a model can be tuned for a device and used to select super-row and super-super-rows sizes in constant time.
\end{abstract}
\end{frontmatter}

\section{Introduction}
Sparse matrix-vector multiplication (\spmv) has remained one of the most important kernels in high-performance computing (HPC) due to its use in applications such as machine learning and iterative solvers (e.g., Conjugate Gradient (CG) and GMRES for partial differential equations (PDEs))~\cite{padma}.
Modern systems are composed of numerous heterogeneous compute units, e.g., many-core CPUs, GPUs, and FPGAs.
Each compute unit has different microarchitecture properties that influence how they access and process data.
Due to \spmv's importance and shift in system design, a need exists for a heterogeneous format, i.e., a data structure that can be efficiently utilized by different compute units, for \spmv.  
This \spmv format needs to exist in a manner that requires only minor adjustments that can be done quickly when switching between compute units.
Overall, the problem of constructing such a format is difficult.
Even on many-core systems, optimizing the performance of \spmv requires tuning given the sparse matrix, microarchitecture of the hardware, and the number of \spmv operations the application will execute~\cite{oski,csr5,csrk}.
Optimizing on accelerators, such as GPUs, can be even more difficult and many of the optimization choices utilized for many-core systems, such as blocking and base format, are completely different.
In this work, we propose extending the well-known many-core system \csr~\cite{csrk,csrki} format to be heterogeneous for both parallel many-core systems and GPUs as this format is both flexible enough and compatible with many already standard interfaces that utilize CSR format.

\spmv, i.e., $Ax=y$ where $A$ is a sparse coefficient matrix and $x,y$ are dense vectors, is well-known to be a notoriously difficult kernel to optimize because of its memory bandwidth requirements~\cite{spmvhard, spmvhard2, csr5}.
The primary challenge in optimizing \spmv is its relatively poor performance on devices due to low computational intensity, i.e., the number of floating-point operations per memory load is low.
Recalling the general algorithm for \spmv, for each row ($i$) of $A$ the nonzero elements are multiplied by their corresponding elements in $x$ and added together for a single element in $y$ ($y_i$).
Each of these nonzeros in a row results in three loads (i.e., $a_{i,j}$, $x_j$, and $y_i$) and one store must occur per multiply-add operation.
Additionally, there exists little data reuse as each $a_{i,j}$ is only used once and $y_i$ is only used for each row.
Making matters even more difficult, the one array that does see reuse (i.e., $x$) might be accessed effectively at random, depending on the structure of the matrix.
This makes it extremely difficult for cache structures to consistently hold relevant data, and places \spmv firmly on the bandwidth-limited roofline of the roofline model~\cite{roofline}.
As such, \spmv may only see a small fraction of the peak performance of a compute device because a very proportionally large amount of time is spent waiting for memory accesses~\cite{peakperf}.

Currently, \spmv formats such as coordinate list (COO) and compressed sparse row (CSR)~\cite{spd, sparsekit} are used to reduce the memory overhead of storing sparse data.
This reduction aids large sparse matrices to better fit into the main memory and may reduce the number of loads needed to keep data in higher cache levels.
However, they are not designed in consideration of parallel processing and the complex microarchitecture of modern devices, such as nonuniform memory accesses (NUMA) on many-core systems.
A second challenge is the complexity of the optimization. 
This complexity comes in the form of the time to optimize and overhead in the size of the data structure.
For example, the process of autotuning \spmv was popular in the early 2000s~\cite{poski,oski}, and the cost for tuning \spmv on many-core systems by constructing complex multilevel block structures could be amortized over the length of the application.
However, both the computational device and the number of concurrent threads could theoretically change per iteration of the application in modern heterogeneous systems, and the cost of autotuning would be too high on such systems.

Therefore in this paper, we extend \csr for heterogeneous systems.
\csr has shown in the past to be a very efficient format on early-generation multicore systems and is relatively easy to understand and store as it is an extension to the highly used CSR format.
Due to being a direct extension of the CSR format, the values can be accessed by other packages that require CSR format without converting once the sparse matrix has been placed in \csr format.
Moreover, the tuning time is far smaller than other methods (e.g., pOSKI~\cite{oski,poski}) as the parameter space is much smaller, and, as we observe in the paper, the tuning can be done in constant time after training a simple model. 
We note that the goal of \csr is for a performance portable heterogeneous format.
Therefore, the performance goal is to be on par with the state of the art and not necessarily better for every device and sparse matrix structure.

This paper provides the extension of \csr for heterogeneous systems, in particular CPUs and GPUs, in the following manner:
\begin{itemize}
    \item An overview of the implementation changes needed for \csr on \nvidia GPUs
    \item The first model-driven selection of tuning parameters for the \csr format to provide constant-time parameters
    \item A comparison of \csr to \kokkos~\cite{kokkos}, \nvidia's \cusparse, CSR5~\cite{csr5}, and TileSpMV~\cite{tilespmv} on Volta and Ampere 
    \item A comparison of \csr to Intel MKL and CSR5~\cite{csr5} on Xeon Platinum 8380 and Epyc 7742
    \item A demonstration of the ability of \csr to offer portable performance on regular matrices and limitations of \csr with irregular matrices. 
\end{itemize}

\section{Background and Related Work}
\label{sec:background}
In this section, we present some of the most common implementation formats for \spmv on CPU and CPU/GPU heterogeneous formats. 
Through this presentation, we outline the strengths and weaknesses of each and their current state.

\subsection{Formats for CPU}
Though many \spmv formats have been studied~\cite{spd, sparsekit,vuduc,bcsr1,ubcsr,csrk},
two standard formats have become popular for \spmv due to reducing memory overheads and their relative ease of understanding, i.e., Coordinate list (COO) and Compressed Sparse Row (CSR).
COO is the most straightforward format for storing a sparse matrix. 
The COO format contains three arrays (i.e., \texttt{col\_idx}, \texttt{row\_idx}, and \texttt{vals}) each array is the length of the number of nonzeros (NNZ) in the sparse matrix.
However, several drawbacks exist for utilizing COO on many-core systems.
These drawbacks include the overall storage required is $3\times NNZ \times 32$ bits if we consider $32$ bit integers and single-precision floating-point values, and the format itself does not provide any ordering of how the nonzero values should be processed.
As a result, a \spmv of a sparse matrix that is stored in COO based on a random permutation of \texttt{col\_idx} and/or \texttt{row\_idx} would have poor performance, and lock or atomic operations would be needed for parallel implementations. 

\begin{figure}[tbh]
    \centering
    \includegraphics[width=.38\textwidth]{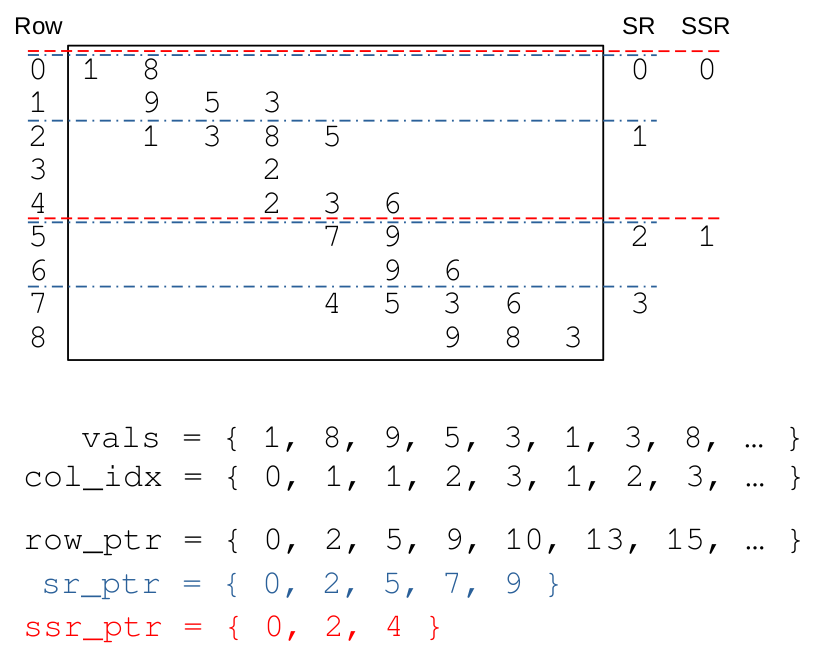}
    \caption{Example of the \csrt data structure with the super-row pointer (\texttt{sr\_ptr}) and super-super-row pointer (\texttt{ssr\_ptr}). Super-rows and super-super-rows are shown in the matrix above. They contain the continuous sums of the number of rows and super-rows, respectively. This is similar to how \texttt{row\_ptr} contains the continuous sum of nonzeros per row.}
    \label{fig:csrk}
\end{figure}

CSR is designed to be a more space-efficient data structure than COO. 
It does this by blocking nonzero elements into shared rows.
The format contains three arrays (i.e., \texttt{row\_ptr}, \texttt{col\_idx}, and \texttt{vals}).
The \texttt{row\_ptr} array provides the running cumulative sum of the nonzero elements in each row and hence has size $m+1$ for an $m \times n$ sparse matrix.
The \texttt{col\_idx} provides the column index and the \texttt{vals} provides the values for each of the nonzero elements.
As such, the CSR requires $(2\times NNZ + m+1)\times 32$ bits of data.
Figure~\ref{fig:csrk} provides an example of this format in black (ignoring \texttt{sr\_ptr} and \texttt{ssr\_ptr}).

Sparse blocked based formats, such as Block Compressed Sparse Row (BCSR)~\cite{bcsr1,vuduc} or Unaligned Block Compressed Sparse Row~\cite{ubcsr}, take the idea of CSR into a second dimension.
Nonzero elements are grouped together in two-dimensional blocks that are normally dense.
Many sparse matrices from finite-element analysis often exhibit this dense block substructure. 
These blocks are next grouped together in some outer blocking structure, such as by row in BCSR.
These can be highly optimized due to the number of different parameters they provide, e.g., the number of rows and columns in a block.
However, the performance of BCSR depends on grouping together nonzero elements into a block, therefore BCSR is not ideal for many sparse matrices.
Though these formats can reduce the memory needed to store the sparse matrix, the true advantage of these formats is having smaller structures and memory access patterns that better fit the hierarchical cache memory structures in many-core systems.
In particular, these small dense blocks can better fit into L1 or L2 caches.

In the application space, \spmv formats are hard to separate from reorderings.
Sparse reorderings permute nonzero elements, and these reorderings can provide better memory accesses~\cite{rcm}, improve iteration counts of sparse iterative solvers~\cite{duff}, and can provide better groupings of nonzeros, such as those taken advantage of by many block based formats and supernodes in sparse direct methods~\cite{davis}.
An example of these orderings includes reorderings that reduce the band around the diagonal, i.e., orderings that pull nonzero elements toward the diagonal, such as spectral orderings~\cite{spec} and Reverse Chuthill-McKee (RCM)~\cite{rcm}.
Therefore, it is almost always the case that some reordering is applied to the sparse matrix in conjunction with selecting a storage format.

\begin{lstlisting}[caption={\csrt CPU kernel}, label=list:serial, frame=single, basicstyle=\footnotesize, morekeywords={function,for,to,let, pragma, omp, parallel}, float, floatplacement=tbh, numbers=left, xleftmargin=2.75em, frame=single, framexleftmargin=2.25em]
function SpMV_3(num_ssr, ssr_ptr[],
                sr_ptr[], r_ptr[], col_idx[],
                vals[], x[], y[]) {
  #pragma omp parallel for            
   for i = 0 to num_ssr {
    let ssr_start = ssr_ptr[i]
    let ssr_end = ssr_ptr[i + 1]
        
    for j = ssr_start to ssr_end {
      let sr_start = sr_ptr[j]
      let sr_end = sr_ptr[j + 1]
            
      for k = sr_start to sr_end {
        let r_start = r_ptr[k]
        let r_end = r_ptr[k + 1]
        let temp = 0.0
            
        for l = r_start to r_end {
          temp += vals[l] * x[col_idx[l]]
        }
        y[k] = temp
} } } }
\end{lstlisting}

\subsection{\csr}
One CSR based format that has shown to greatly improve parallel performance and be more aware of the hierarchical cache structure is \csr~\cite{csrk,csrki}.
\csr is a multilevel data structure for storing sparse matrices based on CSR.
This format utilizes both multilevel structures that better fit the cache structure of many-core systems and multiple levels of reorderings to reduce the envelope or band size for a level.
In particular, the $k$ represents a small integer $\geq2$ that defines the number of additional arrays that store data about rows (i.e., the number of additional arrays equals $k$-1). 
We explore the choice of $k$ as a parameter in Section~\ref{sec:k}.
Figure~\ref{fig:csrk} provides an example of \csrt.
As noted before, the standard three arrays of CSR (i.e., \texttt{vals}, \texttt{col\_idx}, and \texttt{row\_ptr}) exist in \csrt.
CSR-$3$ has two additional arrays.
These arrays are \texttt{sr\_ptr} and \texttt{ssr\_ptr} that contain pointers to the super-rows and the super-super-rows.
Super-rows are groups of contiguous rows, and the array is compressed in a similar manner to how \texttt{row\_ptr} compresses information on the number of contiguous nonzeros in a row. 
As an example from Figure~\ref{fig:csrk}, the SR 0 contains two rows ($2$), the SR 1 contains three rows ($2+3$), the SR 2 contains two rows ($2+3+2$), and finally SR 3 contains two rows ($2+3+2+2$) to provide \texttt{sr\_ptr = \{0,2,5,7,9\}}. 
This method is extended up to super-super-rows that group together contiguous super-rows, i.e., SSR 0 contains two SRs and SSR 1 contains two SRs to provide \texttt{ssr\_ptr = \{0,2,4\}}.
As a result, the \csr format's only memory overhead is the storage of these additional arrays.
Moreover, applications and libraries that have not been fitted to utilize \csr, such as visualization tools or checkpointing, can still process the format as they would a normal CSR format without the overhead of storing both.

For completeness, we provide the pseudocode for \spmv with CSR-3 on a many-core system in Listing~\ref{list:serial}.
Lines 4-5 parallelize the outermost \texttt{for} loop, which iterates over super-super rows (e.g., \texttt{num\_ssr} = 3 in Fig~\ref{fig:csrk}).
In lines 6-9, the bounds for one super-super-row are fetched and iterated across (e.g., \texttt{ssr\_ptr} in Fig~\ref{fig:csrk}).
Similarly, lines 10-13 fetch the bounds for one super-row and are iterated across (e.g., \texttt{sr\_prt} in Fig~\ref{fig:csrk}). 
Finally, lines 14-21 implement the actual work in a similar manner to a CSR-based kernel.

\begin{figure*}[tb!]
\centering
\subfloat[Original]{\label{fig:mo}
\includegraphics[width=.15\textwidth]{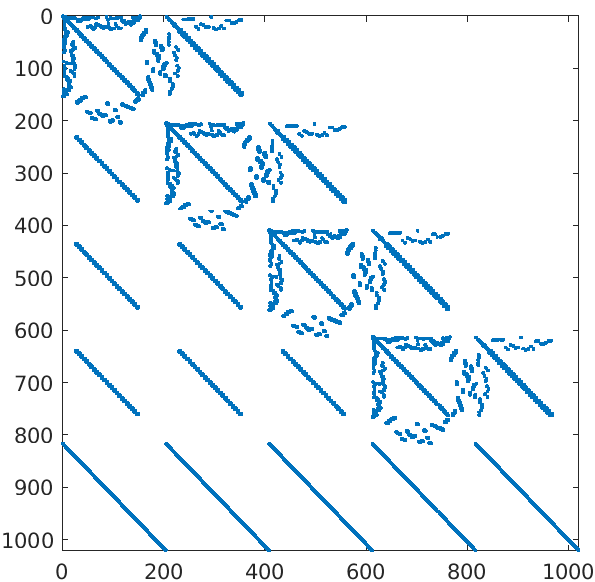}}
\subfloat[Coarsened]{\label{fig:mc}
\includegraphics[width=.15\textwidth]{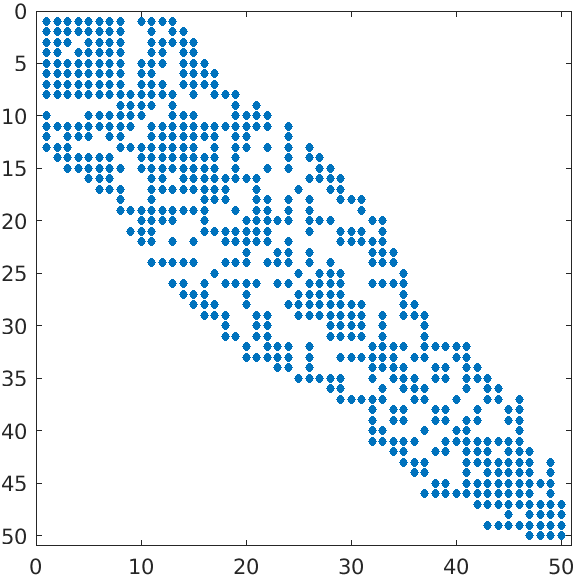} } 
\subfloat[Coarsened Reordered]{\label{fig:mco}
\includegraphics[width=.15\textwidth]{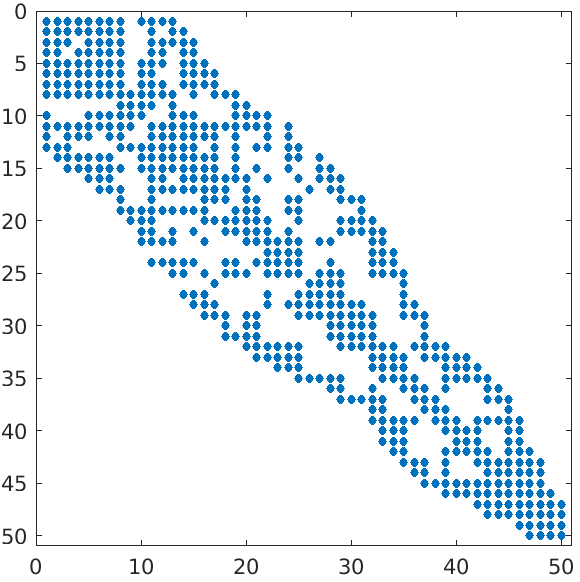}}
\subfloat[Total Reordered]{\label{fig:mt}
\includegraphics[width=.16\textwidth]{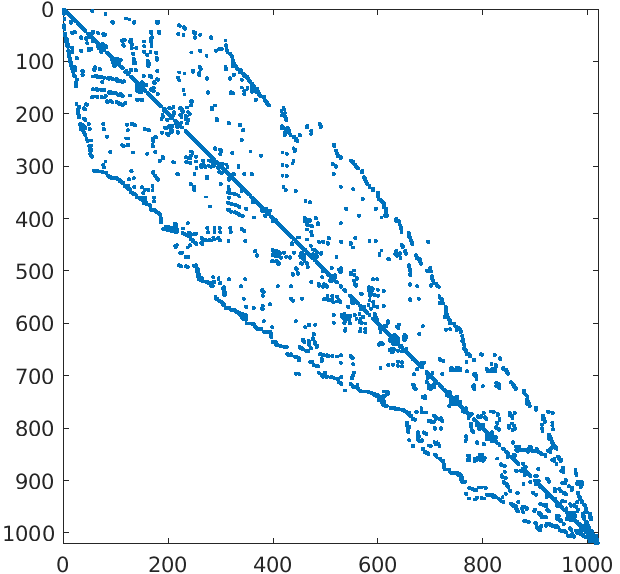} } 
\caption{Band-$k$ algorithm applied to \texttt{circuit204}.}
\label{fig:bandkfig}
\end{figure*}

\begin{lstlisting}[caption={Band-$k$},
label=list:bandk, frame=single, basicstyle=\footnotesize,
morekeywords={function,for,to,let, pragma, omp, parallel}, float, floatplacement=H, numbers=left, xleftmargin=2.75em, frame=single, framexleftmargin=2.25em, mathescape=true]
function BandK($G_{0}$,k) {
   for i = 0 to k - 1 {
      From $G_{i+1}$ from $G_{i}$ using graph coarsening
      Reorder $G_{i+1}$ using  
         weighted bandwidth limiting ordering
   }
   for i = k to 2 {
      Expand node $G_{i}$ to $G_{i-1}$
      for v $\in G_{i}$ {
      Reorder its corresponding vertices
         in $G_{i-1}$ with weighted bandwidth
         limiting ordering
      }
   }
   for i = 1 to k {
      Use reordering of $G_{i}$ to determine 
         super-rows of $A$
}  }
\end{lstlisting}

\csr supports multiple orderings.
In particular, one for \spmv and one for sparse triangular solve as both require very different access patterns and have different structures (i.e., the ordering for \spmv tries to reduce access distance and the ordering for sparse triangular solve tries to minimize dependencies in the elimination graph).
The ordering used by \csr for \spmv is called Band-$k$.
Figure~\ref{fig:bandkfig} provides an example of the Band-$k$ being applied to \texttt{circuit204} from the SuiteSparse collection.
Visually, the ending reordering looks similar to any other bandwidth limiting ordering (e.g., RCM) but the multiple levels, which are not normally visible by the human eye, map better to the \csr format.
For the sake of completeness, we provide a general overview of the Band-$k$ algorithm below and in Listing~\ref{list:bandk}.
The sparse matrix is first converted into a graph $G$.
The graph is coarsened in multiple levels determined by $k$ (i.e., lines 2-6, Fig~\ref{fig:mc}). 
Each coarsened level is reordered using a weighted bandwidth limiting ordering (i.e., lines 7-14, Fig.~\ref{fig:mco}).
Additionally, nodes are reordered within each of these coarsened nodes using a bandwidth limiting ordering.
This process normally produces an ordering with a band slightly wider than that produced by RCM, but better fits the format structure (i.e., Fig~\ref{fig:mt}).
As such, we utilize the standard Band-$k$ format for our extension as well. 
Moreover, the multiple levels of the coarsened-band ordering aim to help with load imbalance. 

\subsection{Formats for GPU}
One notable format has a long history of being pushed on GPU, namely ELLPACK (ELL) format~\cite{ell} and its derivatives.
ELL format is a blocking format, which takes an $m\times n$ sparse matrix and stores it as two $m\times k$ dense matrices, where $k$ is the number of nonzeros in the densest row of the sparse matrix. 
One matrix stores the nonzeros shifted left, padded on the right with zeros. The other matrix stores their corresponding columns also padded with zeros. ELL and its many variants were used heavily for GPU-based SpMV in the early days of GPGPU, due to its friendliness to vector architectures~\cite{ellrt,adell}, but suffer from the issue of excess memory overhead. For instance, if an irregular sparse matrix had the densest row containing 40 nonzeros, but an average nonzero count of 10, the ELL format would incur a 300\% memory overhead.

\subsection{Heterogeneous Formats}
Heterogeneous formats are those that are designed to fit multiple different computational devices.
These have become popular due to the importance of heterogeneous computing systems.
Several popular formats stand out in this area.
The first is the use of blocked sparse format on GPU devices.
Eberhadt and Hoemmen~\cite{bcsr1} demonstrate that BCSR is a reasonable format for both many-core CPUs and GPUs when the sparse matrix contains some block substructure.
However, a specialized algorithm for \spmv on GPU must be reworked for the particular device such as their \emph{by-column} algorithm in order to out-compete \nvidia's built-in sparse matrix library \cusparse.
On the other hand, Liu and Vinter~\cite{csr5}'s \csrf introduces a CSR based format that utilizes a number of tiles to improve performance on both many-core CPUs and GPUs.  
These tiles are of size  $\sigma$ by $\omega$ and are autotuned to better fit the SIMD nature of the device.
The elements in the \texttt{col\_idx} are ordered in a way to fit these tiles and additional vectors (i.e., a title pointer and a title descriptor with lengths that are based on the choice of $\sigma$ and $\omega$) are kept for title information along with a bit-flag vector the length of the number of nonzeros in the sparse matrix.
Therefore, the format is tuned to the device, and storing the spare matrix for additional devices would require keeping additional title pointers and title descriptors. 
Utilizing this format, \csrf can obtain on average the maximum performance of either standard CSR or the autotuned pOSKI~\cite{poski} on many-core CPU systems and is able to obtain on average the maximum performance of all tests kernels on GPUs.
Lastly, work by Aliaga et al.~\cite{ccoo} examines the use of a compressed COO format.
They demonstrate a truly heterogeneous format that can be stored in the same format for both many-core CPUs and GPUs.
This format's performance on CPU is on average better than Intel MKL's CSR utilizing 16 AMD Epyc cores and is better on \nvidia V100 and A100 than \cusparse's COO.
This format compresses values into integer values using a look-up table and requires bit-wise computation to determine blocking.

\section{\csr for Exploiting Multidimensional Block Structure in \nvidia GPUs}
In this section, we provide an overview of how to extend \csr for \nvidia GPUs utilizing CUDA.
We note that there is nothing different in the storage format of \csrt for a GPU and the storage format of \csrt for a CPU that is introduced in the previous section.
The difference is how we interpret the hierarchical levels within \csr and how this optimally gets implemented in the GPU algorithm.
As noted before, the multiple levels in the \csr for a CPU can be interpreted as blocking sets of rows into super-super-rows and super-rows that better fit the hierarchical level cache structure on many-core systems.
The algorithm implementation of this is straightforward with each level providing an additional level of for-loops over the super-super-rows and super-rows.
Listing~\ref{list:serial} provides an example of this code for \csrt on a CPU.

\begin{figure}[tbh]
\centering
\includegraphics[width=.34\textwidth]{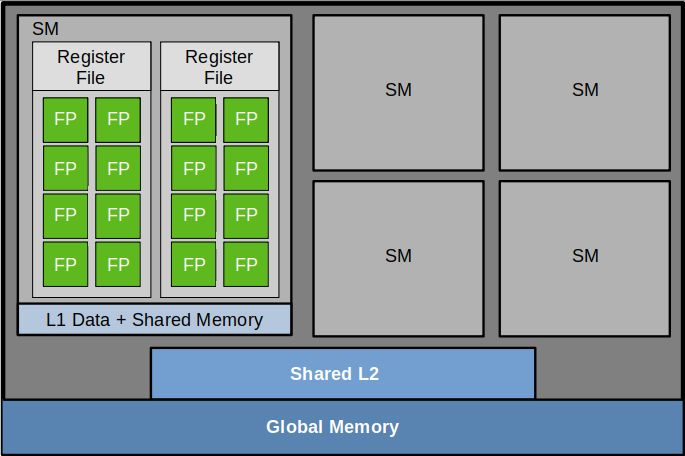}
\caption{Diagram of the \nvidia GPU memory hierarchy. A global shared memory (e.g., HBM2E) is at the bottom. A more standard L2 is shared by streaming multiprocessors (SM).  Each SM has its own L1 that can act as a private data cache or shared memory. Each FP is a floating-point unit.}
\label{fig:gpustruct}
\end{figure}

However, \nvidia GPUs utilizing \cuda do not have the same types of hierarchical level cache structure as a CPU.
Figure~\ref{fig:gpustruct} provides a representation of the memory and execution structure of a \nvidia GPU.
The GPU has a large (e.g., 80 GB) shared global main memory consisting of HBM2 or HBM2E high-bandwidth memory for higher end systems, a shared L2 cache, and a memory layer that can be partitioned as a private L1 data cache or shared memory for each streaming multiprocessor (SM). 
We also note that these caches (such as the L1 data cache) are much smaller per thread than a traditional CPU cache.
For instance, AMD Rome (e.g., AMD Gen2 Epyc) has 32 KB per core of private L1 data cache, while the \nvidia Ampere A100 has 192 KB per SM of L1 data cache. Since each SM has 64 32-bit floating-point units (FP), that equates to an average of 3 KB/thread, though this cache is shared across all FP execution units in an SM.
Additionally, how these resources are scheduled is fundamentally different than a CPU.
Each thread on a CPU can be fixed or pinned to a particular computational core and remain there until it has completed its work.
In the \csr case, this work could be the execution of a particular set of super-rows.
However, 32 threads are blocked together to form a warp on these GPUs.
It is more convenient to think of this warp as a single SIMD intrinsic on CPUs.
Groups of these threads/warps combine to form a thread block, and a thread block supports at most 1024 threads.
These thread blocks are assigned to an SM and will not migrate to other SMs (i.e., they are fixed or pinned).
Therefore, it is easier to think of each block as an assignment of a core to a super-row.
This mapping only goes so far as each SM may have multiple thread blocks and the order of execution of these thread blocks is unknown.
These blocks are further coarsened into a grid.
Due to how the scheduler is implemented, the most efficient \cuda kernels are mapped into a computational grid or cube. 
This allows better exploitation of the 3D nature of \cuda blocks and the grid. 
Utilizing this structure often aids in better data locality, since the \cuda runtime is able to schedule threads close in a block closer on the die. 
Using this cubic environment, we can map a multilevel data structure efficiently to a 2D or 3D block.

\begin{figure}
\centering
\includegraphics[width=.38\textwidth]{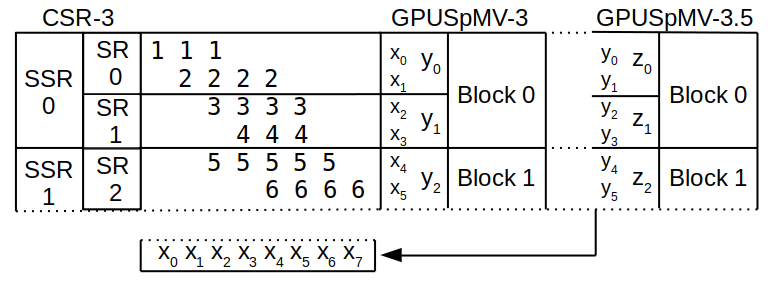}
\caption{CUDA grid and block layout for GPU \spmv algorithms GPUSpMV-3 and GPUSpMV-3.5 in relationship to CSR-3. Note that CUDA views computation and access in the blocks and the three dimensions of $x,y, z$. For GPUSpMV-3, we assign each SSR to a block, each SR to a $y$, and each row to an $x$. We do not use the $z$ dimension. For GPUSpMV-3.5, we assign SSR to a block, each SR to a $z$, each row to a $y$, and each element in a row to a $z$. }
\label{fig:spmvtruct}
\end{figure}

Therefore, we can utilize the hierarchical structure of the GPU even if not directly related to the cache level size for \csr, and in the next section, we justify super-super-row and super-row sizes based on this argument if we utilize \csrt format.
As such, we will primarily focus on a $k =3$ for these \nvidia GPUs.
We start by assuming that an SM is a core or shared-memory computational area in a standard CPU with a shared memory cache, i.e., the L1 data cache, even though each SM may have multiple blocks mapped to it and contains multiple dispatch units.
This rationality is reasonable as a core may work on multiple super-super-rows and these super-super-rows should be issued as a chunk.
Therefore, we can view a single block as a super-super-row each being assigned to an SM.
As such, we build a grid of these blocks, i.e., a 1D grid.
We note that a single block can only support up to 1024 threads, which will put a cap on the maximum size of our super-super-rows.
Within a block, the next level (i.e., super-row) loop can be mapped to the $y$-dimension and the row can be mapped to the $x$-dimension to form 2D blocks.
We note that the order that $x$- and $y$-dimensions are structured in the code is important for GPU dispatch. Warps are grouped first by thread adjacency in the block $x$-dimension, then the $y$-dimension, then the $z$-dimension. Therefore, organizing the threads that work on the inner loops (which work on spatially adjacent data) along the $x$-dimension first is important for data locality.
We call this base code GPUSpMV-3.

Figure~\ref{fig:spmvtruct} provides a diagram of the CSR-3 format's layout in terms of \cuda thread blocks.
As such, super-super-row 0 (SSR 0) will map onto Block 0.
The $y$-dimension of the block will be mapped on the super-row (SR), e.g., $y_0$ maps to SR0 and $y_1$ maps to SR1.
The accompanying pseudocode is provided in Listing~\ref{list:cuspmv3}.
We note that this code is relatively simple and easy to implement in \cuda as it relies on the simple CSR-3 structure and \cuda block parallel functions.

The last for-loop, i.e., the inner product of the sparse matrix row and the vector, (Line 19-21 in Listing~\ref{list:cuspmv3}) in the algorithm provides an additional level of possible parallelism from across the row (i.e., both the multiply-add operations and the sum reduction). 
The original \csr paper did not consider this level of parallelism because of overheads with aligning and data movement for SIMD intrinsics in the test hardware.
As noted in the last section, \csrf explicitly lays out its tiles for these SIMD operations but at the cost of the complexity of the format. 
This level of parallelism is important and needs to be addressed in GPU codes.
However, if the number of nonzeros per row is relatively small (e.g., $< 8$), utilizing this level of parallelism would reduce the available number of overall threads to a block and would require the overhead of utilizing part of the L1 data cache for shared memory.
Through experimentation, we discovered that 8 nonzero elements per row is what is required to improve performance with parallelization at this level.
We denote the algorithm with the parallelism of the last level as GPUSpMV-3.5, and pseudocode is provided in Listing~\ref{list:cuspmv35}. Figure~\ref{fig:spmvtruct} also provides the structure for GPUSpMV-3.5. 
However, super-super-rows are now the blocks, the super-rows are the $z$-dimension, the rows are the $y$-dimension, and nonzeros in the rows are the $x$-dimension due to CUDA convention.
For optimization purposes, we provide a \texttt{temp} array in \cuspmvtf is allocated as a block-local shared memory space for fast read and write access.

\begin{lstlisting}[caption={\cuspmvt GPU kernel}, label=list:cuspmv3, frame=single, basicstyle=\footnotesize, morekeywords={function,for,to,let,parallelize,across,no,parallelization,__global__}, float, floatplacement=tbh, numbers=left, xleftmargin=2.75em, frame=single, framexleftmargin=2.25em]
function cuSpMV_3(num_coarsest_rows, ssr_ptr[],
                  sr_ptr[], r_ptr[], col_idx[],
                  vals[], x[], y[]) {
  let block = blockIdx.x
  let ssr_start = ssr_ptr[block]
  let sr_end = ssr_ptr[block + 1]
        
  for i = ssr_start to ssr_end  
    parallelize across blockDim.y {
    let sr_start = sr_ptr[i]
    let sr_end = sr_ptr[i + 1]
            
    for j = sr_start to sr_end
      parallelize across blockDim.x {
      let r_start = r_ptr[j]
      let r_end = r_ptr[j + 1]
      let temp = 0.0
            
      for k = r_start to r_end
        no parallelization {
        temp += vals[k] * x[col_idx[k]]
      }
                
      y[k] = temp
} } }
\end{lstlisting}

\section{Tuning Optimal Structure}
\label{sec:tuning}
\subsection{GPU}
When tuning \csr for CPUs normally a low cost autotuning method is needed, and the original paper notes how inexpensive this autotuning is compared to pOSKI.
Autotuning on a GPU opens up many new parameters, such as block dimensions and if the inner product should be parallelized (e.g., GPUSpMV-3 vs GPUSpMV-3.5).
However, there exist some standards that can help guide this tuning, and these standards can even be reduced to a closed form heuristic formula that can determine the super-row and super-super-row sizes in constant time for a given sparse matrix after some initial tuning on a given device.

\begin{lstlisting}[caption={\cuspmvtf GPU kernel}, label=list:cuspmv35, frame=single, basicstyle=\footnotesize, morekeywords={function,for,to,let,parallelize,across,no,parallelization,__global__,parallel_reduction,fill}, float, floatplacement=tbh, numbers=left, xleftmargin=2.75em, frame=single, framexleftmargin=2.25em]
function cuSpMV_3.5(num_coarsest_rows, ssr_ptr[],
                    sr_ptr[], r_ptr[], col_idx[],
                    vals[], x[], y[]) {
  let block = blockIdx.x
  let ssr_start = ssr_ptr[block]
  let ssr_end = ssr_ptr[block + 1]
        
  for i = ssr_start to ssr_end 
    parallelize across blockDim.z {
    let sr_start = sr_ptr[i]
    let sr_end = sr_ptr[i + 1]
            
    for j = sr_start to sr_end
      parallelize across blockDim.y {
      let r_start = r_ptr[j]
      let r_end = r_ptr[j + 1]
      let temp[blockDim.x] = fill(0.0)
            
      for k = r_start to r_end
        parallelize across blockDim.x {
        temp[threadIdx.x] +=
          vals[k] * x[col_idx[k]]
      }
      y[k] = parallel_reduction(temp)
} } }
\end{lstlisting}

The first of these standards is based on the relationship between the number of threads and block size.
Remember that only 1024 threads can be used per block and that 32 threads are launched at a time in a warp.
Therefore, the dimensions of the block should be based on of multiples of 32.
The second of these standards is having enough work to keep each thread busy while limiting the working memory size of each thread so accesses are not hitting slower caches or main memory.
Additionally, making more threads is more likely to amortize any imbalance in work each thread might have. 
The average amount of work a single row must compute (in \cuspmvt) is based on the average row density, i.e., $rdensity = NNZ/N$ where $NNZ$ is the number of nonzeros and $N$ is the number of rows.
Additionally, we have experimentally determined as noted before that serial computation of the inner product per row only makes sense when $rdensity < 8$.
Based on these two facts and common block dimensions used in GPUs, we only need to consider the following cases:

\vspace{.25cm}
\noindent \textbf{Case 1: $rdensity \leq 8$} \\
The block dimensions are set as $8\times 12$.\vspace{.25cm}

\noindent \textbf{Case 2: $8 < rdensity \leq 16$} \\
The block dimensions are set as $4\times 8\times 12$.\vspace{.25cm}

\noindent \textbf{Case 3: $16 < rdensity \leq 32$} \\
The block dimensions are set as $8\times 8\times 8$.\vspace{.25cm}

\noindent \textbf{Case 4: $32 < rdensity$} \\
The block dimensions are set as $16 \times 8\times 4$.\vspace{.25cm}

\noindent We can interpret these values as follows.
In Case 1, four rows along the $y$-dimension (i.e., $8\times 4$ threads) would make up a warp.  
Since the $y$-dimension corresponds to a super-row in \cuspmvt, a single warp would have four super-rows.
This may seem like a lot of work, but the $rdensity$ is relatively small in this case.
On the other hand, in Case 4, two rows along the $y$-dimension would make up a warp due to the number of nonzeros in a single row needing to be computed.

Once these standard block sizes are set, empirical runs can be completed with different sizes of super-rows and super-super-rows over a suite of sparse matrices.
We can use these runs with the following modeling method to determine a closed form heuristic for the selection of super-super-row and super-row sizes in the future.
Using these test runs, we perform a logarithmic regression over the dataset, with the $x$-values being $rdensity$ and the $y$-values being the optimal super-super-row or super-row sizes.
The $SSRS$ (super-super-row size) and $SRS$ (super-row size) parameters are tuned independently.
That is, during testing, all reasonable combinations of $SSRS$ and $SRS$ are tested (some set of representative sizes between $4$ and $48$ for GPU) and the regression is performed twice: one on the optimal $SSRS$ parameters, and one on the optimal $SRS$ parameters. 
The specific set of combinations for $SSRS$ and $SRS$ tested are described as follows:
$$\left(SSRS,SRS\right)\in\left(\bigcup_{i=2}^{5}\left\{2^i,1.5\cdot 2^i\right\}\right)^2$$
where $\phi^2$ is shorthand for the set Cartesian product $\phi\times\phi$.

Finally, since the logarithmic regression tends to yield a formula that drops much below optimal when $rdensity$ becomes large, the coefficient of the natural logarithm was lowered by hand to better fit the optimal $SSRS$ and $SRS$ with high $rdensity$.
 
From this method, we achieve the following formula for tuning on the \nvidia Volta:
$$ SSRS=\nint*{8.900-1.25\cdot\ln{(rdensity)}} \mbox{   and}$$
$$ SRS=\nint*{10.146-1.50\cdot\ln{(rdensity)}}$$
where $\nint*{x}$ represents rounding-to-nearest, half towards positive infinity operation.

These numbers would have to be derived for each different machine that has considerable microarchitecture differences.
For \nvidia Ampere, a slight tuning is required and provides: 
$$ SSRS=\nint*{9.175-1.32\cdot\ln{(rdensity)}} \mbox{   and}$$
$$ SRS=\nint*{20.500-3.50\cdot\ln{(rdensity)}}.$$

Similarly, the coefficient of the natural logarithm is lowered so that  $SSRS$ and $SRS$ do not drop too low on higher density rows.
The constant term is unchanged in both instances.

After the initial $SSRS$ and $SRS$ are determined, they can be further tuned based on $rdensity$. 
In particular, the $SSRS$ and $SRS$ need to be updated to account for their derivation from the original assumption of utilizing GPUSpMV-3 to GPUSpMV-3.5, in addition to accounting for the changing block dimensions as $rdensity$ grows.
As such, they are modified on Volta:
\vspace{.1cm}

\noindent \textbf{Case 1: $rdensity \leq 8$} \\
Tune SSRS and SRS no further.\vspace{.25cm}

\noindent \textbf{Case 2: $8 < rdensity \leq 16$} \\
$SSRS = \nint*{SSRS\times 1.5} $\\
$SRS = SRSS \times 2$.\vspace{.25cm}

\noindent \textbf{Case 3: $16<rdensity \leq 32$} \\
$SSRS = SSRS\times 4$\\
$SRS = \floor{SSRS / 2} $.\vspace{.25cm}

\noindent \textbf{Case 4: $32<rdensity$} \\
$SSRS = SSRS\times 5$\\
$SRS = \floor{SSRS / 2}$.\vspace{.25cm}

For Ampere, they are modified as:
\vspace{.1cm}

\noindent \textbf{Case 1: $rdensity \leq 8$} \\
Tune SSRS and SRS no further.\vspace{.25cm}

\noindent \textbf{Case 2: $8 < rdensity \leq 16$} \\
Tune $SSRS$ no further.\\
$SRS = SRSS \times 4$.\vspace{.25cm}

\noindent \textbf{Case 3: $16<rdensity \leq 32$} \\
$SSRS = \nint*{SSRS\times 2.5}$\\
$SRS = SSRS \times 3 $.\vspace{.25cm}

\noindent \textbf{Case 4: $32<rdensity$} \\
$SSRS = SSRS \times 2$\\
$SRS = SSRS \times 2$.\vspace{.25cm}

\subsection{CPU}
\label{sec:cputune}
When running CSR-$2$ on a CPU, the kernel is much less dependent on having optimal $SSRS$ and $SRS$ within a narrow window for good performance.
Additionally, the added complexity of the cache hierarchy in CPUs, as opposed to GPUs, makes tuning much more difficult to do optimally, as there is no clear correlation between optimal $SSRS$/$SRS$ and known matrix attributes without analyzing matrix structure.
Due to these combinations of factors, we find that in an ideal scenario, each matrix would be tuned individually, using a set of representative $SRS$ between 8 and 3072, which are specified as the set:
$$SRS\in\bigcup_{i=3}^{11}\left\{2^i,1.5\cdot 2^i\right\}$$
The combined effect of larger caches on CPU and using CSR-$2$ instead of CSR-$3$ means that much higher $SRS$ are often preferred.
Namely, most matrices on CPU using CSR-2 prefer $SRS$ in the range of 40-1000, as opposed to most matrices on GPU using \csrt preferring $SSRS$/$SRS$ in the range of 4-12.

If constant-time tuning is preferred at the expense of performance, we have found that taking the geometric mean of optimal $SRS$ across a representative dataset yields decent performance. For instance, one might use a $SRS=96$ for all matrices on CPUs to yield performance that is often ``good enough.''

\section{Test Setup}
\label{sec:testsetup}

\subsection{Test Systems}

\begin{table}[]
\caption{Benchmark suite. ID identifies regular vs. irregular, SY is the percentage of pattern symmetry, N is the outer dimension in millions, NZ is the number of nonzeros in millions, MAX is the maximum number of nonzeros per row, and R is the row density. }
\label{tab:testsuite}
{\footnotesize
\begin{tabular}{|l|l|l|l|l|l|l|} \hline
ID  & SY  & Matrix                   & N (M) & NZ (M) & MAX    & R     \\ \hline
r1  & 1.0 & belgium\_osm             & 1.44  & 3.10   & 10     & 2.15  \\
r2  & 1.0 & netherlands\_osm         & 2.22  & 4.88   & 7      & 2.20  \\
r3  & 1.0 & roadNet-TX               & 1.39  & 3.84   & 12     & 2.76  \\
r4  & 1.0 & roadNet-CA               & 1.97  & 5.53   & 12     & 2.81  \\
r5  & 1.0 & roadNet-PA               & 1.09  & 3.08   & 9      & 2.83  \\
r6  & 1.0 & hugetrace-00000          & 4.59  & 13.76  & 3      & 3.00  \\
r7  & 1.0 & hugetric-00000           & 5.82  & 17.47  & 3      & 3.00  \\
r8  & 1.0 & debr                     & 1.05  & 4.19   & 4      & 4.00  \\
r9  & 1.0 & venturiLevel3            & 4.03  & 16.11  & 6      & 4.00  \\
r10 & 1.0 & rajat31                  & 4.69  & 20.32  & 1.2K   & 4.33  \\
r11 & 1.0 & G3\_circuit              & 1.59  & 7.66   & 6      & 4.83  \\
r12 & 1.0 & ecology1                 & 1.00  & 5.00   & 5      & 5.00  \\
r13 & 1.0 & 333SP                    & 3.71  & 22.22  & 28     & 5.98  \\
r14 & 1.0 & AS365                    & 3.80  & 22.74  & 14     & 5.98  \\
r15 & 1.0 & NACA0015                 & 1.04  & 6.23   & 10     & 5.99  \\
r16 & 1.0 & M6                       & 3.50  & 21.00  & 10     & 6.00  \\
r17 & 1.0 & NLR                      & 4.16  & 24.98  & 20     & 6.00  \\
r18 & 1.0 & delaunay\_n20            & 1.05  & 6.29   & 23     & 6.00  \\
r19 & 1.0 & delaunay\_n21            & 2.10  & 12.58  & 23     & 6.00  \\
r20 & 1.0 & delaunay\_n22            & 4.19  & 25.17  & 23     & 6.00  \\
r21 & 1.0 & atmosmodl                & 1.49  & 10.32  & 7      & 6.93  \\
r22 & 1.0 & atmosmodm                & 1.49  & 10.32  & 7      & 6.93  \\
r23 & 1.0 & atmosmodd                & 1.27  & 8.81   & 7      & 6.94  \\
r24 & 1.0 & atmosmodj                & 1.27  & 8.81   & 7      & 6.94  \\
r25 & 1.0 & thermal2                 & 1.28  & 8.58   & 11     & 6.99  \\
r26 & 1.0 & CurlCurl\_3              & 1.22  & 13.54  & 13     & 11.11 \\
r27 & 1.0 & CurlCurl\_4              & 2.38  & 26.52  & 13     & 11.14 \\
r28 & 1.0 & StocF-1465               & 1.47  & 21.01  & 189    & 14.34 \\
r29 & 1.0 & Transport                & 1.60  & 23.48  & 15     & 14.66 \\
r30 & 1.0 & packing...               & 2.15  & 34.98  & 18     & 16.30 \\ 
r31 & 1.0 & af\_shell10              & 1.51  & 52.26  & 35     & 34.65 \\ \cline{2-7}
r32 & 0.0 & Hamrle3                  & 1.45  & 5.51   & 9      & 3.81  \\
r33 & 0.9 & circuit5M\_dc            & 3.52  & 14.87  & 24     & 4.22  \\
r34 & 0.9 & memchip                  & 2.71  & 13.34  & 27     & 4.93  \\
r35 & 0.9 & Freescale1               & 3.43  & 17.05  & 25     & 4.97  \\ \hline
i1  & 1.0 & com-Youtube              & 1.13  & 5.98   & 28.8k  & 5.27  \\
i2  & 1.0 & kkt\_power               & 2.06  & 12.77  & 90     & 6.19  \\
i3  & 1.0 & FullChip                 & 2.99  & 26.62  & 2.3M   & 8.91  \\
i4  & 1.0 & circuit5M                & 5.56  & 59.52  & 1.3M   & 10.71 \\
i5  & 1.0 & as-Skitter               & 1.70  & 22.19  & 35.5K  & 13.08 \\
i6  & 1.0 & rgg\_n\_2\_20\_s0        & 1.05  & 13.78  & 36     & 13.14 \\
i7  & 1.0 & rgg\_n\_2\_21\_s0        & 2.10  & 28.98  & 37     & 13.82 \\
i8  & 1.0 & cage14                   & 1.51  & 27.13  & 41     & 18.02 \\
i9  & 1.0 & dgreen                   & 1.20  & 26.61  & 97     & 22.16 \\
i10 & 1.0 & nv2                      & 1.45  & 37.48  & 84     & 25.78 \\
i11 & 1.0 & nlpkkt80                 & 1.06  & 28.19  & 28     & 26.54 \\
i12 & 1.0 & Hook\_1498               & 1.50  & 59.37  & 93     & 39.64 \\
i13 & 1.0 & dielFilterV2real         & 1.16  & 48.54  & 110    & 41.94 \\ \cline{2-7}
i14 & 0.1 & wiki-Talk                & 2.39  & 5.02   & 3.3K   & 2.10  \\
i15 & 0.2 & wiki-talk-temporal       & 1.14  & 3.31   & 3.3K   & 2.90  \\
i16 & 0.1 & webbase-1M               & 1.00  & 3.11   & 28.6K  & 3.11  \\
i17 & 0.0 & patents                  & 3.77  & 14.97  & 762    & 3.97  \\
i18 & 0.0 & cit-Patents              & 3.77  & 16.52  & 779    & 4.38  \\
i19 & 1.0 & Freescale2               & 3.00  & 14.31  & 14.9K  & 4.77  \\ 
i20 & 0.1 & wikipedia-2005110       & 1.63  & 19.75  & 75.5K  & 12.08 \\
i21 & 0.4 & in-2004                  & 1.38  & 16.92  & 21.9K  & 12.23 \\
i22 & 0.1 & wikipedia-2006092       & 2.98  & 37.27  & 159K & 12.49 \\
i23 & 0.1 & wikipedia-2006110       & 3.15  & 39.38  & 168K & 12.51 \\
i24 & 0.1 & wikipedia-2007020       & 3.57  & 45.03  & 187K & 12.62 \\
i25 & 0.4 & sx-stackoverflow         & 2.60  & 36.23  & 22.3K  & 13.93 \\
i26 & 0.2 & wiki-topcats             & 1.79  & 28.51  & 238K   & 15.92 \\
i27 & 0.5 & soc-Pokec                & 1.63  & 30.62  & 230K   & 18.75 \\
i28 & 0.7 & ss                       & 1.65  & 34.75  & 27     & 21.03 \\
i29 & 0.4 & vas\_stokes\_1M          & 1.09  & 34.77  & 1K     & 31.88 \\\hline
\end{tabular}
}
\end{table}

The test setup consists of four systems. 
\textbf{GPU Systems.}
System 1 is used for GPU testing and has two Xeon E5-2650v4 CPUs (``Broadwell'') with 12 cores each, and contains several \nvidia V100 GPUs (``Volta'').
These GPUs contain 32 GB of main memory and a peak memory bandwidth of 900 GB/s.
System 2 is also used for GPU testing and has two Epyc 7713 CPUs (``Milan'' or ``Zen 3'') with 64 cores each, and contains \nvidia A100 GPUs (``Ampere'').
These GPUs contain 40 GB of memory and a peak memory bandwidth of 1555 GB/s.
On both GPU systems, only one GPU is used.
\textbf{CPU Systems.}
System 3 is for CPU testing. 
It contains two Epyc 7742 CPUs (``Rome'' or ``Zen 2'') and 256 GB of memory in 8 memory channels.
Simultaneous multi-threading (SMT) is disabled for system 3.
System 4 is also for CPU testing and contains two Intel Xeon Platinum 8380 CPUs (``Ice Lake Server'') and 256 GB of memory.
SMT is enabled for system 4, but we opt not to use it in tests.

\subsection{Tested Libraries}
For GPU experiments, \cuspmvt and \cuspmvtf are both implemented in C++ with \cuda and compiled with the \nvidia NVHPC compiler using compilation tools version 11.4. 
Our \csr \spmv implementations are compared against \cusparse v11.4, \kokkos v3.4.1~\cite{kokkos}, \csrf~\cite{csr5}, and \tile~\cite{tilespmv}.
\nvidia's \cusparse is an optimized sparse linear algebra library written for use on \nvidia GPUs. 
It provides \spmv implementations in several formats, and we compare it against the library's CSR implementation.
\nvidia's \cusparse is part of the \cuda runtime, so all runs are performed using \cuda runtime 11.4, which has support for both Volta and Ampere GPUs.
\kokkos is a computational mathematics library published by Sandia National Laboratories.
It supports sparse and dense linear algebra, among other kernels, on both CPUs and GPUs.
The goal of \kokkos is to provide performance portable code to multiple parallel runtime libraries without requiring the user to code in the different libraries. \kokkos is compiled for the SM\_70 compute architecture, which is supported by our Volta GPUs. 
The tested version of \kokkos does not support building for multiple GPU architectures (e.g., SM\_70 and SM\_80), and therefore will only be tested on Volta. 

\csrf is tested on both Volta and Ampere and is compiled with the same compiler used for \cuspmv.
We were unable to use the \texttt{-gencode} flag the default make has to specify the GPU architecture because the implementation uses several functions that are deprecated for SM\_70 and onwards, which includes both of our GPU test beds.
\tile is tested on Ampere and is compiled with the same compiler used for \cuspmv.
\tile is a format for sparse matrices that stores tiles of the matrix in one of seven formats: \csrf, ELL, HYB, COO, dense column, dense row, and dense.
A decision tree classifies each tile into one of the storage formats.
At runtime, optimized device kernels are chosen on a per-tile basis to match the storage format of a given tile.
It is important to note that \tile is not advertised as a heterogeneous solution.
As such, it is not tested on CPUs.
We do not test it on Volta, as we use it for a replacement for \kokkos.

For CPU experiments, we utilize Intel MKL v19.1.3 on system 3 and Intel MKL v19.1.1 on system 4.
Intel MKL (Math Kernel Library) is a computational mathematics library that supports highly-optimized sparse linear algebra on CPUs.
\csr is compiled on system 3 with the AMD Optimizing C Compiler (AOCC) v2.3.0 with \texttt{-O3} optimization and \texttt{-march=znver2}.
\csr is compiled on system 4 with the Intel v19.1.1 compiler with \texttt{-O3 -ipo} optimization and \texttt{-march=icelake-server}.
Additionally, OpenMP scheduling parameters are set to \texttt{static} scheduling for \csr on systems 3 and 4.

\csrf is also tested on systems 3 and 4, using the same compiler configurations as \csr.
We compile the AVX2 version of \csrf for both systems 3 and 4.
Despite system 4 supporting AVX512, the AVX512 implementation of \csrf uses extensions that are not supported in our Ice Lake CPU (namely, \texttt{AVX512ER} and \texttt{AVX512PF}).

\subsection{Test Suite}
For our test suite, we consider all sparse square matrices with the outer dimension ranging from 1 to 6 million and having less than 60 million nonzeros from the SuiteSparse collection~\cite{suitesparse}.
This provides a total of 64 sparse matrices with various sparsity patterns.
Table~\ref{tab:testsuite} provides a list of these sparse matrices along with a categorization (ID), percentage of the symmetric pattern (SY), outer dimension (N) given in millions, number of nonzeros (NZ) given in millions, the maximum number of nonzeros in a row (MAX), and the average number of nonzeros in a row (i.e., row density) (R).
Matrices within the table are categorized into two levels of groups: regular vs. irregular and symmetric vs. nonsymmetric.
The group of regular vs. irregular is similar to the grouping done in the analysis of CSR5~\cite{csr5}, and provides a way to judge if the number of nonzeros in a row varies from the average.
As such, sparse matrices whose number of nonzeros do not vary greatly from the average are considered regular, while sparse matrices whose number of nonzeros vary greatly from the average are considered irregular.
While the analysis of CSR5 used the maximum and the minimum number of nonzeros in a row to approximate this grouping into regular vs irregular, we calculated the variance of the number of nonzero in a row  ($\sigma^{2}$) and used the value of $\sigma^2 = 10$ to be the point of classifying into the two groups of regular and irregular.
Next, we classify the groups of regular and irregular into symmetric and nonsymmetric.  
Any matrix that does not have a 100\% (1.0) symmetric sparsity pattern is considered nonsymmetric. 
While this classification may not be interesting to formats like CSR5, we believe that this is an interesting factor for us as our reordering algorithm considers the symmetrical pattern of the matrix during the coarsening and reordering phases.  
Lastly, matrices within the two levels are sorted based on their average number of nonzeros per row.

Additionally, we want to note that matrices from different application areas tend to be placed within a certain category of regular vs. irregular and symmetric vs. nonsymmetric.
For example, sparse matrices from the solution of large PDEs utilizing finite-difference or finite-element methods such as those from computational fluid dynamics (\texttt{atmosmod}, \texttt{StocF-1465}), triangulations (\texttt{delaunay}), or thermal problems (\texttt{termal2}) tend to be symmetric in pattern and regular.
On the other hand, matrices from areas such as circuit simulations (\texttt{Harmle3}, \texttt{memchip}, \texttt{FullChip}) or graph analysis (\texttt{wiki-Talk}, \texttt{sx-stackoverflow}) tend to be more nonsymmetric and irregular.

For \kokkos, \cusparse, and MKL, all sparse matrices are first reordered utilizing RCM obtained from MATLAB's \texttt{symrcm}.
As noted before, RCM provides improve convergence for some iterative solvers (e.g., CG) and improves performance due to reducing the stride of irregular accesses.
We opt to feed \csr with matrices in their natural ordering to demonstrate the efficacy of the Band-$k$ algorithm~\cite{duff}.
\csrf and \tile are fed matrices in their natural ordering, as the original papers do not specify the use of bandwidth-reducing reorderings.
The rationale for this choice is twofold.
The first is that \csrf and \tile are complex enough formats that already look for clusterings of nonzeros for tiles and therefore impose some ordering themselves. 
This is unlike \cusparse, \kokkos, and MKL, which use simple formats that can benefit from reordering.
The second is that our Band-$k$ algorithm is part of our \csr algorithm and we feed \csr the same natural ordered matrices, therefore it would be a better comparison as we are directly comparing how we form block structures via super-rows and them using their tiling. 

\subsection{Test Methodology}
For GPU tests, data is copied ahead of time to the GPU, and only the \spmv kernel execution time is measured.
This is done to accurately model the behavior of iterative solvers, which should not re-copy the data each iteration.
For both CPU and GPU tests, 20 runs are performed and the results are averaged via arithmetic mean.
On CPU tests, 5 untimed warmup runs are performed, because MKL appears to take 1-2 iterations before reaching maximum performance.

\section{CPU Performance and $k$ Selection}
In this section, we verify the ability of \csr to improve performance on CPUs.
This is necessary as the original paper was printed over eight years ago (2014), and the design of CPUs has changed.
In particular, the three design changes have been increased core count on a single socket, different caching structures (e.g., multiple segmented shared caches as in Rome), and different hardware reuse techniques (e.g., newer Intel systems).
Additionally, we wish to re-investigate the impact of the selection of the $k$ for \csr.
In keeping with the format of other heterogeneous formats presentation, we present the performance as the GFlop/s.

\subsubsection{CPU Performance}
Figure~\ref{fig:cpu} presents the four bar graphs with performance measured in GFlop/s.
The top two bar graphs (\ref{fig:icereg} and ~\ref{fig:iceirreg}) present the performance on the Intel Ice Lake system for the regular and irregular groups.
We divided the symmetric and nonsymmetric sparse matrices within the bar graph with a vertical black dashed bar, and we present the average performance for each method within the category (i.e., within regular or irregular including both symmetric and nonsymmetric) as a horizontal dash line.
This system is most like the one the original \csr work was presented on.
For the regular category, we find that CSR-2 provides the highest average GFlop/s at $\sim 44.1$ with CSR-3 and CSR5 closely behind at $\sim 38.5$ and $\sim 35.3$ (i.e., $\sim 14.5 \%$ and $\sim 24.9\%$ improvement).
Overall CSR-2 does better in this group except in a couple of cases such as: 1.) the first five with an average row density of less than three; 2.) the very nonsymmetric \texttt{Hamrle3}; 3.) \texttt{debr}.
In the case of the irregular sparse matrices on Ice Lake, the gap between CSR-2 and \csrf closes with an average of $\sim 32.2$ GFlop/s and $\sim 28.8$ GFlop/s, respectively (i.e., $\sim 11.8\%$ improvement).
In general, we notice that \csrf is better than CSR-2 in very irregular matrices and those with nonsymmetric structures, such as those from the Wikipedia grouping and \texttt{webbase-1M}.

We next consider the performance on the Rome system (\ref{fig:romereg} and ~\ref{fig:romeirreg}).
This system is of importance due to how different it is from the Intel system in terms of cache size and structure.
In particular, Rome is set up as a cluster of 16 core complexes (CCXs), each consisting of 4 cores that share an L3 cache.
The latency for accessing the L3 cache from a core's own CCX is very low (i.e., around 39 clock cycles) while accessing a remote L3 can be slower than the main memory.
Additionally, Rome has less L1 data cache and L2 cache per core while having over 4$\times$ as much L3 cache~\cite{memperf}. 
We notice that the gap between the performance of CSR-2 and CSR5 is more pronounced on this system.
For the category of regular matrices, CSR-2 can obtain an average performance of $\sim 73.1$ GFlop/s while CSR5 only obtains $\sim 37.2$ GFlop/s (i.e., $\sim 95.7\%$ improvement).   
Moreover, the only two sparse matrices in this group that CSR5 can outperform CSR-2 on are \texttt{debr} and \texttt{Hamrle3}.
We additionally observe better behavior of CSR-2 for the category of irregular matrices on Rome.
CSR-2 is able to obtain an average performance of $\sim 36.5$ GFlop/s while CSR5 is able to obtain $\sim 22.1$ GFlop/s (i.e., $\sim 65.1\%$ improvement)
As with the Intel Ice Lake system, CSR5 does seem to handle irregular and nonsymmetric matrices better than CSR-2.
However, the performance gap is much smaller on the AMD Rome system with CSR-2 outperforming CSR5 in all cases except for the sparse matrix of \texttt{circuit5M}.
Based on these results, we can validate that \csr is still a valid method on CPUs.
We do note that if the matrix is known to be very nonsymmetric or very irregular, CSR5 would be a better choice in particular on the Intel Ice Lake system.

\subsection{Scalability}
\begin{figure}[tbh]
\centering
\subfloat[Scalability on Ice Lake]{\label{fig:sice}
\includegraphics[width=.4\textwidth]{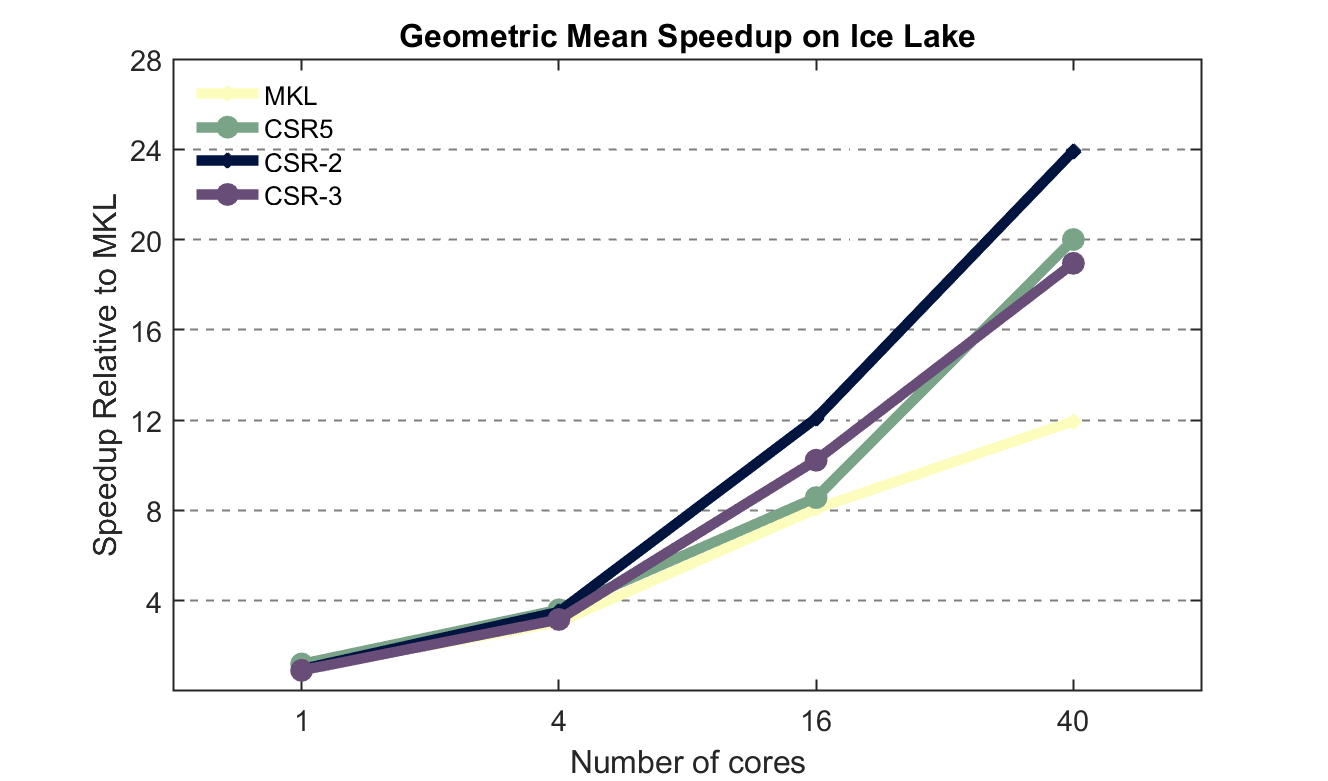}}\\
\subfloat[Scalability on Rome]{\label{fig:srome}
\includegraphics[width=.4\textwidth]{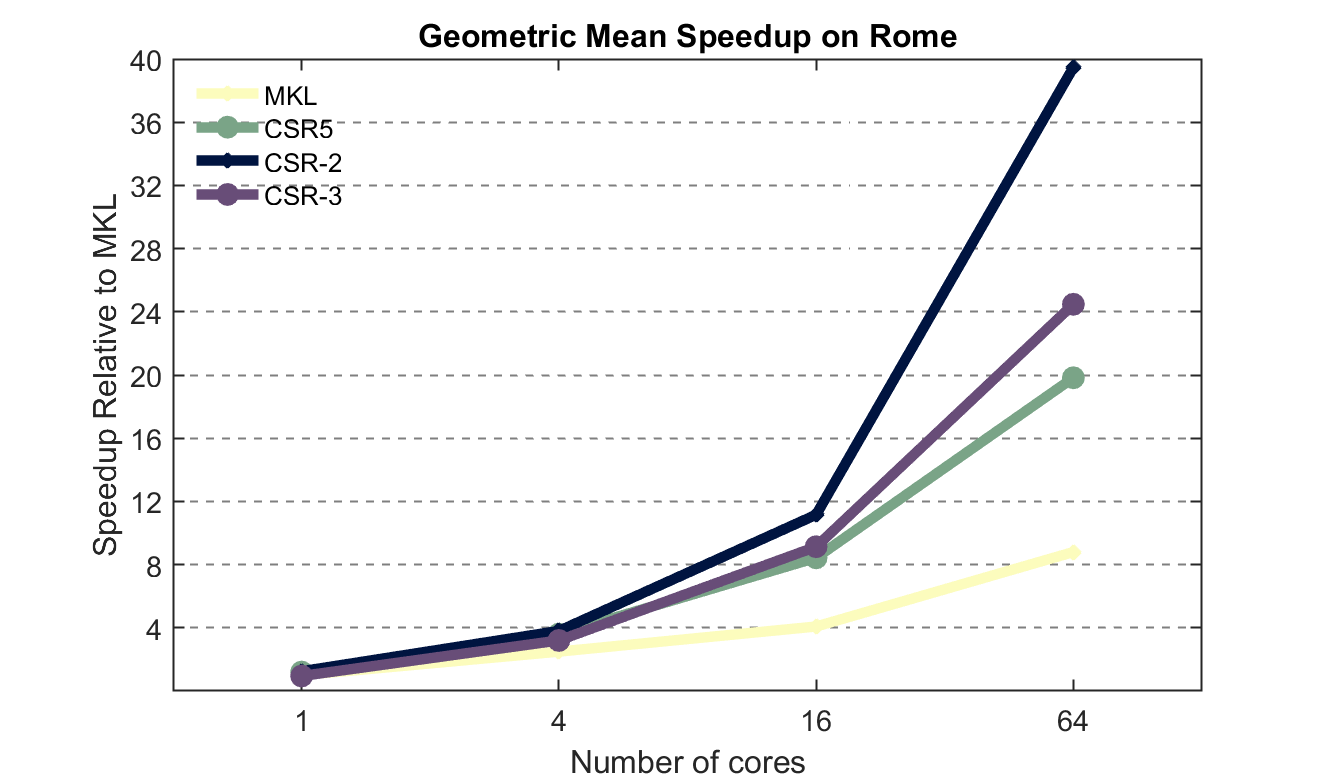}}
\caption{Scalability study.}
\label{fig:scale}
\end{figure}

For completeness, we also present the scalability in terms of speedup relative to MKL for our two CPU systems.
The geometric mean of all 64 speedups (i.e., a speedup for each sparse matrix) is presented.
This analysis helps to determine if there are unforeseen overheads (e.g., the use of parallel data structures or synchronizations) that may plague an implementation at a different number of cores.
Figure~\ref{fig:scale} presents this scalability study.
We observe that on both systems, all SpMV methods are about the same using 4 cores.
On both systems, CSR-2, CSR-3, and CSR5 close (i.e., $\sim 8-12\times$) at 16 cores; however, Intel MKL does not do well on the AMD Rome system.
When utilizing the full chip, the spread of scalability starts to widen with CSR-2 performing better than CSR5.

\subsection{Selection of $k$}
\label{sec:k}
\begin{figure*}[tb!]
\centering
\subfloat[GFlop/s on Ice Lake (40 cores) regular matrices]{\label{fig:icereg}
\includegraphics[width=.98\textwidth]{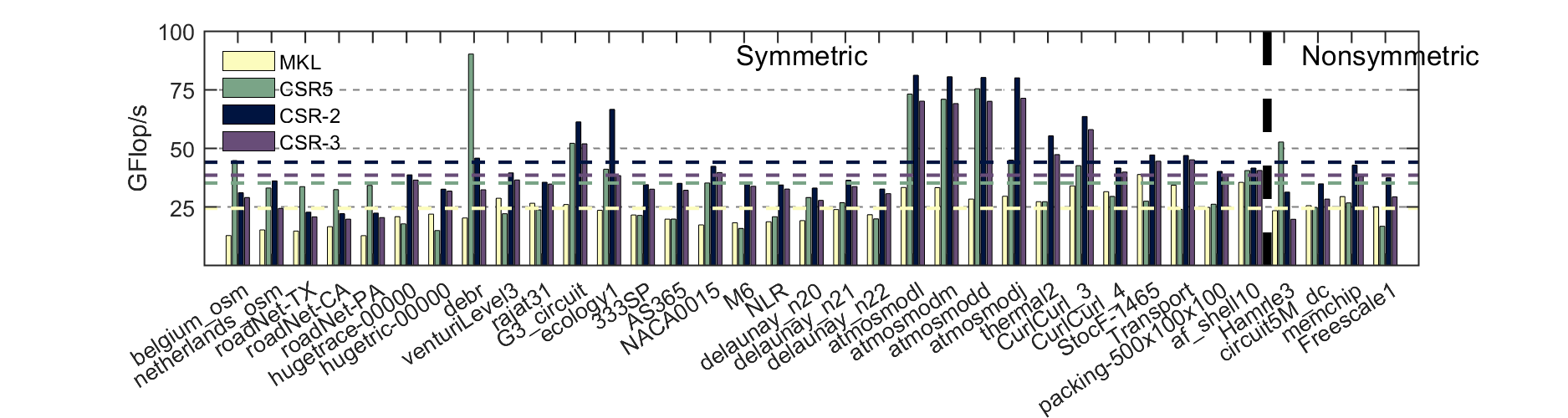}}\\
\subfloat[GFlop/s on Ice Lake (40 cores) irregular matrices]{\label{fig:iceirreg}
\includegraphics[width=.98\textwidth]{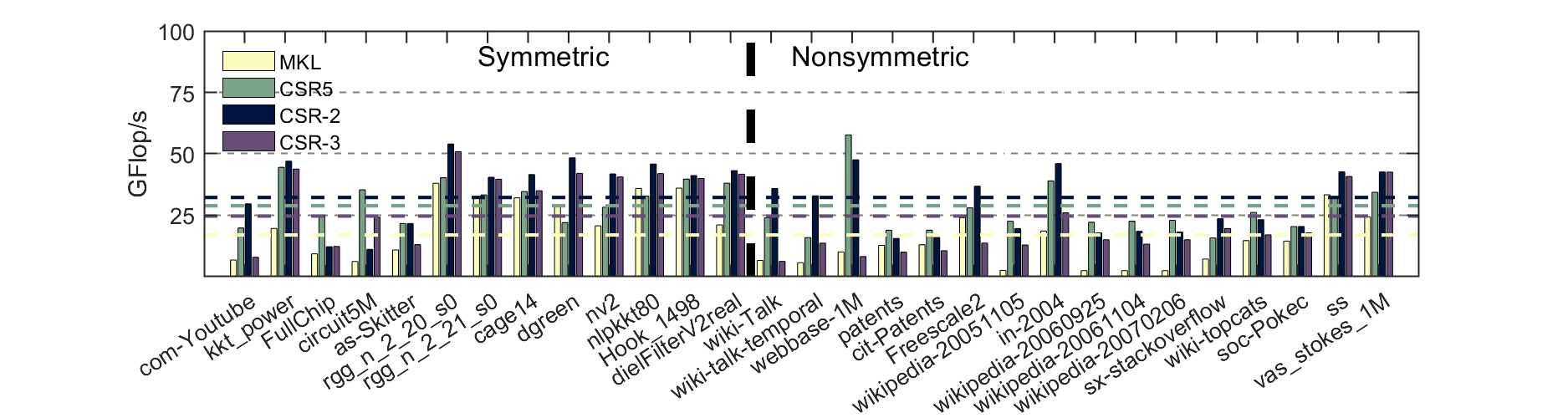} } \\
\subfloat[GFlop/s on Rome (64 cores) regular matrices]{\label{fig:romereg}
\includegraphics[width=.98\textwidth]{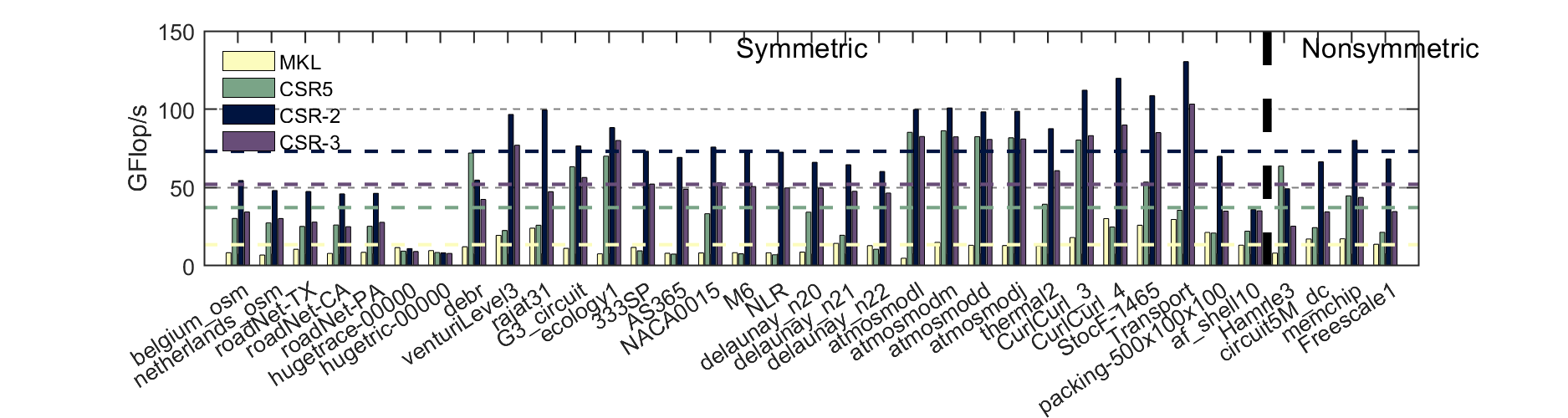}}\\
\subfloat[GFlop/s on Rome (64 cores) irregular matrices]{\label{fig:romeirreg}
\includegraphics[width=.98\textwidth]{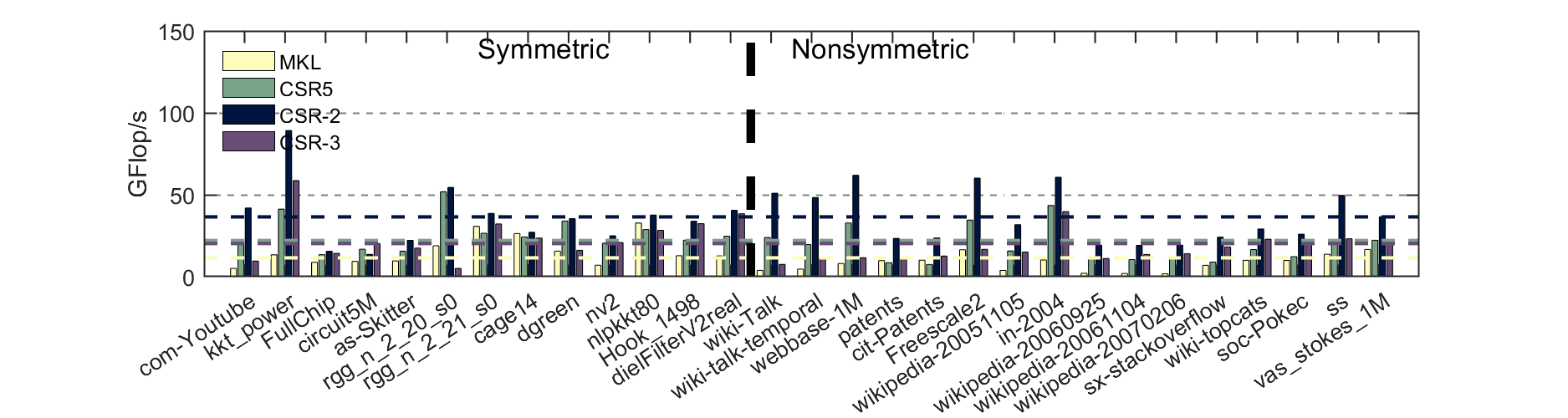} } 
\caption{Performance on CPU}
\label{fig:cpu}
\end{figure*}

Like the validity of the CPU performance of the original work, the selection of $k$ should also be validated.
In the original work, the choice of $k$ is made based on the number of discrete cache levels in the hardware and not on the sparse matrix, which the original work makes based on their own experimental tests.
Therefore, the choice of $k$ for our two systems (Ice Lake and Rome) should be 2, based on the observations of the original work.
We did additional runs with $k = 3$ (i.e., CSR-3) to validate if this choice is something that could be set independently of the sparse matrix.
Figure~\ref{fig:cpu} presents the performance of CSR-3 for both systems and both categories.
From this data, CSR-2 always outperforms CSR-3 on average.
On the Ice Lake system, CSR-3 is never able to outperform CSR-2 even with the additional cost of tuning both the super-row and super-super-row size.
On the Rome system, CSR-3 does seem to perform better but is still not able to outperform CSR-2. 
Therefore, we validate the same observation they made for the CPU systems.

In the next section, we present the GPU performance. 
Based on our algorithm and our view of the CUDA memory system, we likewise predict that CSR-3 will be sufficient.
We note that we also test CSR-2 on the GPU, and we find that in a couple of matrices CSR-2 slightly outperforms CSR-3.
We do not include all of the CSR-2 results for the GPU in the figures due to readability of the bar graphs but do present the listing of matrices outperformed by CSR-2 in the text.
However, the number of sparse matrices that are outperformed by CSR-2 is so few that we would generally recommend CSR-3, and why we presented CSR-3 for GPU as our main algorithm.

\begin{figure*}[tb!]
\centering
\subfloat[GFlop/s on Volta regular matrices]{\label{fig:voltareg}
\includegraphics[width=.98\textwidth]{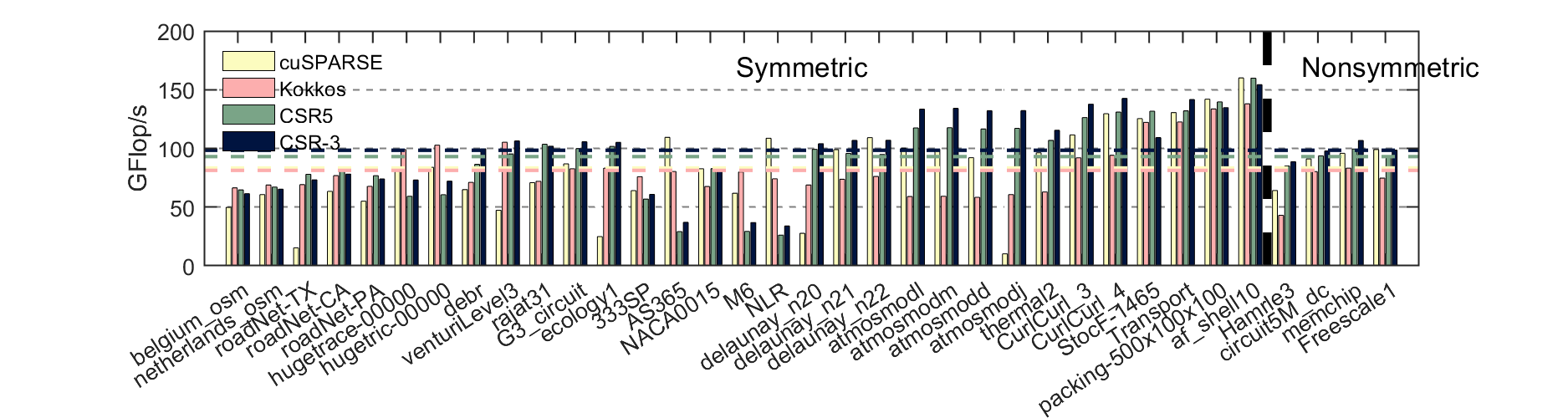}}\\
\subfloat[GFlop/s on Volta irregular matrices]{\label{fig:voltairreg}
\includegraphics[width=.98\textwidth]{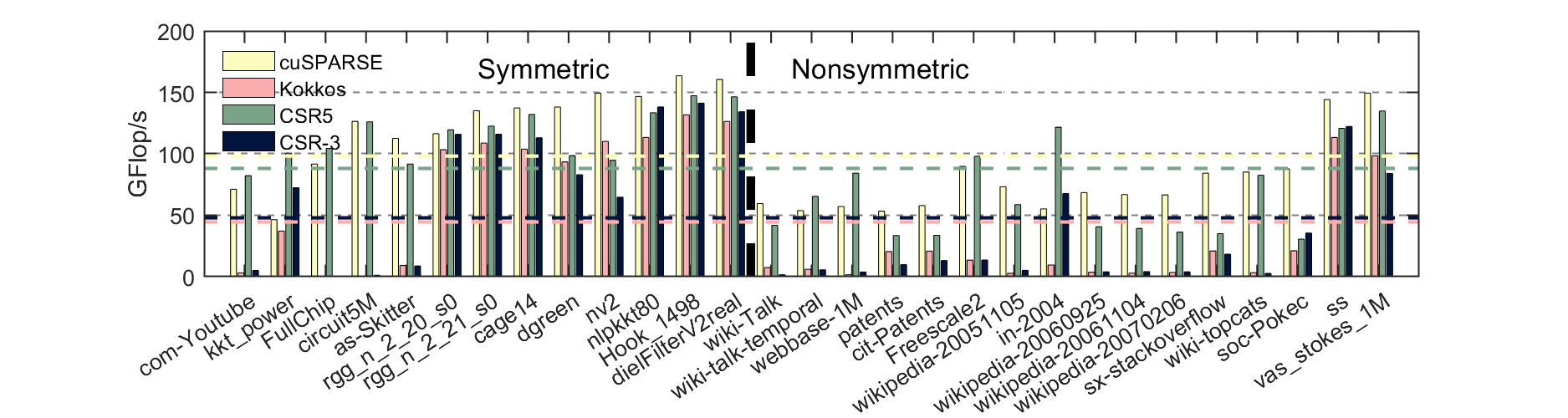} } \\
\subfloat[GFlop/s on Ampere regular matrices]{\label{fig:ampreg}
\includegraphics[width=.98\textwidth]{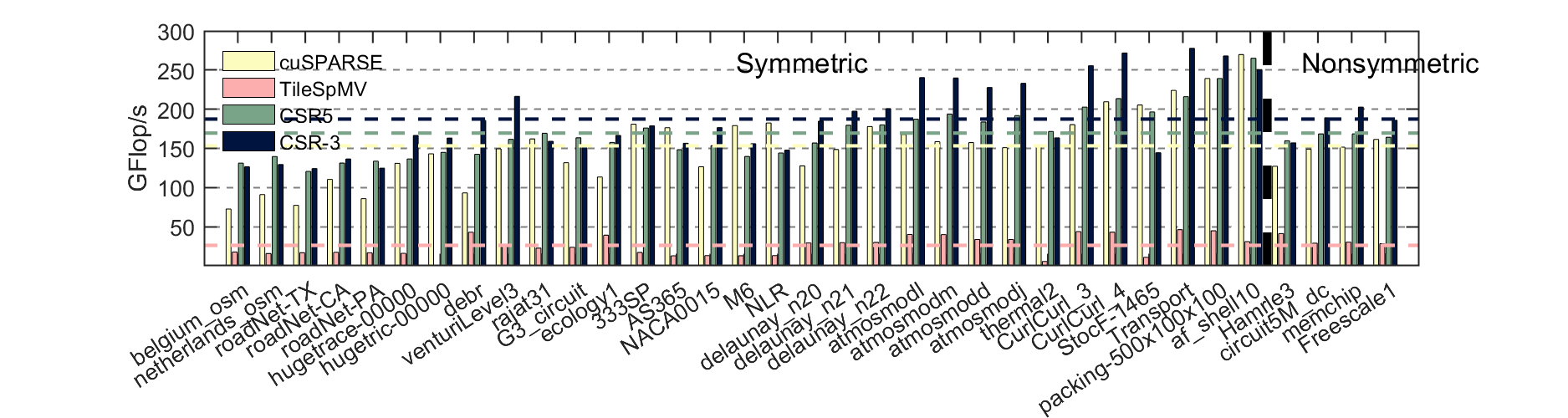}}\\
\subfloat[GFlop/s on Ampere irregular matrices]{\label{fig:ampirreg}
\includegraphics[width=.98\textwidth]{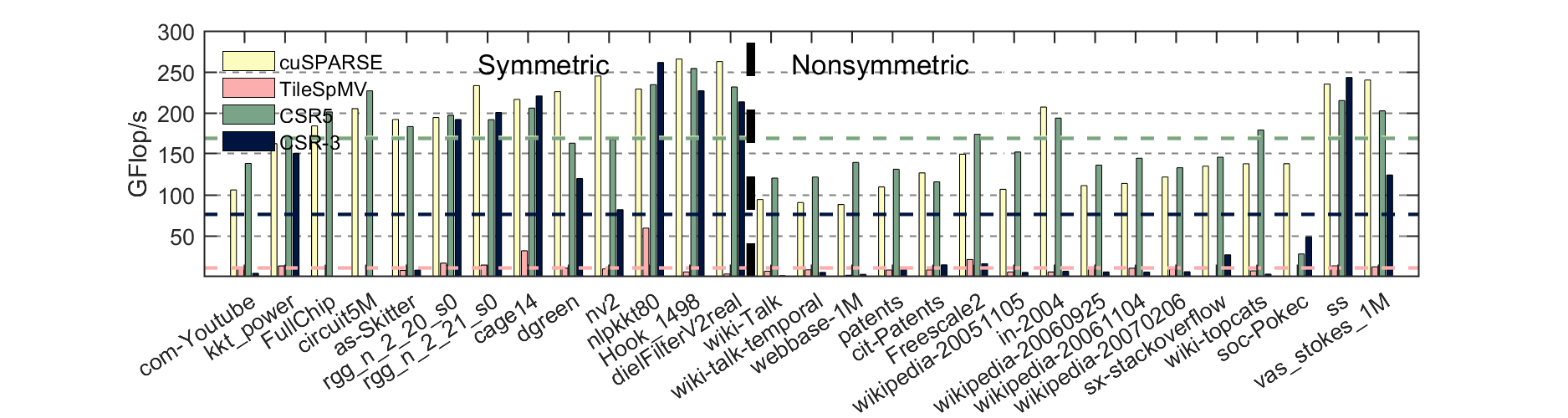} } 
\caption{Performance on GPU}
\label{fig:gpu}
\end{figure*}

\section{GPU Performance}
\label{sec:gpu}
In this section, we analyze the performance of \csrt on Volta and Ampere GPUs.
We tune the super-super-row and the super-row sizes based on the automatic analysis tuning in Section~\ref{sec:tuning}. 
We compare these results to both the CSR representation format in \cusparse and \kokkos for the Volta and only \cusparse on the Ampere due to \kokkos's current compiler limitations. 
We test against the state of the art format of \csrf on both Volta and Ampere as this is a true heterogeneous format.
As a replacement for \kokkos on Ampere, we test on the state of the art (non-heterogeneous) format of \tile. 
Similar to the last section, we utilize GFlop/s as our primary measurement of performance.

Figure~\ref{fig:gpu} presents the performance measurements on GPU.
Results for the regular and irregular categories are present in ~\ref{fig:voltairreg} and ~\ref{fig:voltairreg} for the Volta system.
Again, the average performance for a method within the regular and irregular categories is presented as a dashed line.
In the regular category, CSR-3 achieves an average of $\sim 98.4$ GFlop/s while CSR5 achieves $\sim 93.0$ GFlop/s (i.e., $\sim 5.8\%$ improvement). 
Similar to CSR-2 vs. CSR5 on the CPU system of Ice Lake, CSR5 does better than CSR-3 on the first five sparse matrices with an average row density of less than three. 
We also observe that \kokkos does well in this group and is the fastest on the next two sparse matrices, i.e., \texttt{hugetrace-00000} and \texttt{hugeetric-00000}, while CSR-3 does very poorly on these. 
We believe this is an artifact of how the \kokkos implementation is tuned as the CSR-3, CSR5, and cuSPARSE implementations perform very similarly to each other and \kokkos is the only outlier. 
The choice of the best method is much less cut and dry on Volta than in the CPU systems.
One example of this is \texttt{AS365}, \texttt{M6}, and \texttt{NLR} where both \cusparse and \kokkos outperform CSR-3 and CSR5.
However, the tables are opposite for \texttt{CurlCurl\_3} and \texttt{Hamrle3}.
We also observe that CSR-3 is less sensitive to the nonsymmetric nature of \texttt{Hamrles3} on this system.
Considering the irregular category of sparse matrices on the Volta system, \cusparse ($\sim 98.2$ GFlop/s) and CSR5 ($\sim 88.15$ GFlop/s) outperform CSR-3 ($\sim 47.9$ GFlop/s) on average.
We note that CSR-3 is particularly bad on irregular sparse matrices with a very low average row density (i.e., the first five matrices).
However, we also note that these five matrices have some of the largest maximum nonzero row counts which makes them some of the most irregular matrices.
This is likely due to the fact that the selection between \cuspmvt and \cuspmvtf is based on average row density, so the very dense rows are being computed in serial.
A potential avenue for future work could include selecting based on maximum row density for such highly irregular matrices, which would allow for aforementioned dense rows to be computed in parallel.
As the average row density increases (and likewise the irregularity decreases) in our test suite, CSR-3 starts to perform better.

Additionally, CSR-2 runs are done to compare the sensitivity of the selection of $k$ (not pictured in the figure).
CSR-2 was able to outperform CSR-3 by more than $5\%$ in four cases (\texttt{wiki-Talk}, \texttt{patents}, \texttt{as-Skitter}, and  \texttt{debr} ).
The most improvement of CSR-2 compared to CSR-3 was in the case of \texttt{wiki-talk} by $~25.1\%$. 
However, due to how poorly CSR-3 performs on \texttt{wiki-talk}, CSR-2 is still outperformed by cuSPARSE and CSR5 on this matrix.
Therefore, we conclude that the in the regular case the choice of either CSR-3 or CSR5 makes sense while the choice of either cuSPARSE or CSR5 in the irregular case makes sense.

Ampere is the newer NVIDIA GPU device and the performance of our test suite is presented in ~\ref{fig:ampreg} and ~\ref{fig:ampirreg}.
We note that CSR-3 ($\sim 187.3$ GFlop/s) outperforms CSR5 ($\sim 169.5$ GFlop/s) and cuSPARSE ($\sim 153.3$ GFlop/s) on average by about $10.5\%$ and $22.2\%$, respectively. 
Both CSR-3 and CSR5 do not have the same reduced performance on \texttt{AS365}, \texttt{M6}, and \texttt{NLR} as they did on the Volta system.
Moreover, CSR-3 is not as negatively impacted by the low row density of the first five matrices as it was on the Volta system.
For the irregular category, cuSPASRE ($\sim 169.9$ GFlop/s) outperforms CSR5 ($\sim 168.8$ GFlop/s) and CSR-3 ($\sim 76.1$ GFlop/s) on average by $\sim 0.6\%$ and $\sim 123\%$.
Again, CSR-3 suffers from the same issue as it does on Volta for irregular matrices.

Additionally, CSR-2 runs are done to compare the sensitivity of the selection of $k$ (again, not pictured in the figure).
CSR-2 was able to outperform CSR-3 by more than $5\%$ in fives cases (\texttt{wiki-Talk}, \texttt{patents}, \texttt{as-Skitter}, \texttt{debr}, and \texttt{FullChip} ).
Similar to the previous time, CSR-2 outperforms CSR-3 but does not equate to much since the largest outperformance is on sparse matrices that CSR-3 already performs very poorly on. 
For example, CSR-2 outperforms CSR-3 by $~17\%$ on \texttt{wiki-Talk} but this is still not enough to be close to either cuSPARSE or CSR5.
Therefore, we conclude that CSR-3 be an ideal choice if the user knew the sparse matrix was regular on the Ampere system.

\section{Overall Observations About a Heterogeneous Format}
The primary goal of any heterogeneous format is to have a portable format that scales well across multiple different systems.
In particular, our goal is to have a format that scales across CPUs and GPUs.
As seen in the previous two sections, both CSR5 and \csr are not perfect heterogeneous formats, and almost all have some limitations as we are optimizing over such a large array of systems and sparse structures.
In our observations, we see that CSR5 is limited in its performance on both Ice Lake and Rome CPU systems with CSR-2 outperforming CSR5 by as much as $\sim 95.7\%$ on AMD Rome CPUs for regular matrices and as little as $\sim 11.8\%$ on Intel Ice Lake CPUs for irregular matrices.
On GPUs, CSR-3 can still outperform CSR5 but only by a small margin and only in the category of regular sparse matrices.
In particular, CSR-3 can outperform CSR5 by at most $\sim 10.5\%$ on the Ampere system for regular matrices.
However, CSR-3 is unable to keep up with CSR5 at all in the category of irregular matrices on GPU systems.
Therefore, the right choice of which package to use would be based on the primary use of the package.
If the choice is for a package that will be used on GPU most often and needs to support various matrix structures, then CSR5 is a clear winner.
However, if the choice is for a package within a sparse iterative linear solver designed for PDEs (e.g., CG) that might equally be executed on CPU or GPU then \csr makes sense as it has better performance on the CPU systems and slightly better performance than CSR5 on GPU systems for regular matrices that are common in these types of applications.

\section{Conclusion}
An efficient heterogeneous format for \spmv is required by modern high-performance computing. 
This format needs to be easy to tune and provide portable performance across devices. 
This paper presents \csr as a heterogeneous format solution for \spmv, and in particular, for the use on regular matrices, i.e., sparse matrices where the number of nonzeros per row has a variance $\leq 10$.
This kind of sparse matrix is common in the solution of PDEs and could be used in a \spmv driven sparse iterative solvers such as CG.
This CSR-based format is easy to understand, can be tuned quickly, and can be used as-is by library calls that require a standard CSR format.
Moreover, we present a method to tune CSR-3 in constant time for a particular sparse matrix utilizing an autotuned formula model.
Using tuned CSR-2, the format method outperforms \csrf and Intel MKL on Intel Ice Lake and AMD Rome Systems on average over our whole test suite of regular and irregular matrices.
This improvement can be as high as $\sim 95.7\%$ against \csrf on AMD Rome.
Our CSR-3 version for GPU outperforms \csrf  by at most $\sim 10.5\%$ on the Ampere systems for regular matrices but fails to outperform for irregular matrices on GPU systems.
Therefore, we demonstrated that \csr is an ideal heterogeneous format for many-core CPUs and \nvidia GPUs when dealing with regular matrices that tend to be symmetric or close to symmetric in their sparsity pattern.

\section*{Acknowledgement}
\vspace{-6pt}
This work was made possible in part by support from the Alabama Supercomputer Authority, Sandia National Laboratories, and NSF2044633. This work used the Extreme Science and Engineering Discovery Environment (XSEDE), which is supported by National Science Foundation grant number ACI-1548562.

\bibliographystyle{plainnat}
\bibliography{journal}

\end{document}